\documentclass[conference]{IEEEtran}

\usepackage{amssymb,amsfonts}
\usepackage{textcomp}
\usepackage{xcolor}
\usepackage{url}
\usepackage{hyperref}

\definecolor{yellow}{RGB}{255,255,153}
\definecolor{grey}{RGB}{224,224,224}
\definecolor{green}{RGB}{0,100,0}

\usepackage{listings}
\usepackage{algorithm}
\usepackage{algorithmic}
\usepackage{graphicx}
\usepackage{subfigure} 
\usepackage{fancybox}
\usepackage{xspace}
\usepackage{threeparttable}
\usepackage{multirow}
\usepackage{enumitem}
\usepackage{seqsplit}
\usepackage{pifont}
\usepackage{adjustbox}

\begin{document}
\title{Same App, Different Behaviors: Uncovering Device-specific Behaviors in Android Apps}

\author{
    \IEEEauthorblockN{Zikan Dong\IEEEauthorrefmark{1}, Yanjie Zhao\IEEEauthorrefmark{2}, Tianming Liu\IEEEauthorrefmark{3}, Chao Wang\IEEEauthorrefmark{2}, Guosheng Xu\IEEEauthorrefmark{1}, Guoai Xu\IEEEauthorrefmark{4}, Haoyu Wang\IEEEauthorrefmark{2}}
    \IEEEauthorblockA{\IEEEauthorrefmark{1}Beijing University of Posts and Telecommunications, Beijing, China\\
    \IEEEauthorrefmark{2}Huazhong University of Science and Technology, Wuhan, China\\
    \IEEEauthorrefmark{3}Monash University, Melbourne, Australia\\
    \IEEEauthorrefmark{4}Harbin Institute of Technology, Shenzhen, China
    }
}

\maketitle

\begin{abstract}
The Android ecosystem faces a notable challenge known as fragmentation, which denotes the extensive diversity within the system. 
This issue is mainly related to differences in system versions, device hardware specifications, and customizations introduced by manufacturers.
The growing divergence among devices leads to marked variations in how a given app behaves across diverse devices.
This is referred to as device-specific behaviors.
In this work, we present the first large-scale empirical study of device-specific behaviors in real-world Android apps.
We have designed a three-phase static analysis framework to accurately detect and understand the device-specific behaviors.
Upon employing our tool on a dataset comprising more than 20,000 apps, we detected device-specific behaviors in 2,357 of them.
By examining the distribution of device-specific behaviors, our analysis revealed that apps within the Chinese third-party app market exhibit more relevant behaviors compared to their counterparts in Google Play. Additionally, these behaviors are more likely to feature dominant brands that hold larger market shares.
Reflecting this, we have classified these device-specific behaviors into 29 categories based on implemented functionalities, providing structured insight into these behaviors.
Beyond common behaviors like issue fixes and feature adaptations, we observed 33 aggressive apps, including popular ones with millions of downloads, abusing system properties of customized ROMs to obtain user-unresettable identifiers without requiring permission, substantially impacting user privacy.
Finally, we investigated the origins of device-specific behaviors, revealing significant challenges developers face in implementing them comprehensively.
Our research sheds light on the promising but less touched research direction of device-specific behaviors, benefiting community stakeholders.
\end{abstract}

\section{Introduction}

Android stands out as a highly popular mobile operating system, commanding a market share of up to 71.4\%~\cite{market}. 
The success of Android is closely linked to its open nature, yet this openness has given rise to notable fragmentation challenges emanating from the diverse array of devices and varying Android system versions.
In contrast, the closed nature of the iOS ecosystem facilitates a cohesive user experience, with 17 major releases to date supported on just over 40 smartphone models.
The scenario in the Android ecosystem is notably intricate. 
To elaborate, there exist over 20,000 distinct devices operating on the Android system, featuring diverse hardware configurations. 
Even when restricting the focus to those presently trending, their quantity surpasses that of iOS devices by a considerable margin~\cite{fragmentation}. 
Moreover, despite the Android system having released only 13 major versions thus far, certain prominent device manufacturers (such as Samsung, Huawei, and Xiaomi) extensively modify the Google-led Android Open Source Project (AOSP) to appeal to a broader consumer base. 
During the initial phases of system customization, manufacturers concentrated on enhancing the user interface, evident in the incorporation of ``UI'' in the names of numerous customized systems.
Yet, as technology progressed, manufacturers began to progressively alter system functionalities and introduce new features absent in AOSP. 
These factors contribute to an increasing divergence among different Android devices.
The challenge of fragmentation presents hurdles for developers tasked with ensuring a consistent user experience for their app across a diverse range of devices and system versions. 
Consequently, a single app may display distinct behaviors across various devices.
Our research centers around this phenomenon, which we term \textbf{\textit{device-specific behavior}}.

Previous research~\cite{liu2023automatically,wei2016taming,wei2019pivot} has shed light on device-specific issues in the Android ecosystem, primarily focusing on compatibility challenges.
These efforts have led to significant advancements, including the development of automated tools like FicFinder and Pivot, which identify device-specific compatibility issues by detecting patterns and extracting API-device correlations.
However, beyond \textbf{compatibility issue fix} which focuses on resolving hardware-specific or OS--specific glitches, there exists another crucial category of device-specific behavior, \textbf{manufacturer-introduced feature adaptation}.
Such behaviors capitalize on the novel functionalities integrated by manufacturers into their customized systems, providing developers with the opportunity to attract a wider user base.
While existing works~\cite{wei2016taming,wei2019pivot} have touched upon this aspect, it has not been the main focus of their research.

The research gap becomes especially pertinent with \textbf{privacy-related} device-specific behaviors, a domain yet to be investigated in existing literature. 
When customizing their own systems, manufacturers may inadvertently undermine AOSP's security protection, opening new venues for malicious actors to gain access to sensitive user information unscrupulously. 
Reports~\cite{pdd} have indicated that these customizations have already been abused by certain popular apps.
Specifically, by extensively collecting vulnerabilities in various major customized systems, the reported app, whose user base amounts to over 800 million, achieves privileges escalation across numerous mainstream devices and subsequently gains access to sensitive user data, silently monitoring every move of the user.

Therefore, our research aims to provide a comprehensive understanding of device-specific behaviors in large-scale real-world Android apps, moving beyond the traditional lens of compatibility issues to encompass a broader spectrum of device-specific behaviors.
We devised a static analysis pipeline to facilitate our large-scale study.
Taking Android apps as input, we first perform static data flow analysis to identify an extensive amount of code snippets that are associated with device-specific behaviors, and then provide a categorization of these device-specific behaviors based on the specific functionalities they implement.

Applying our pipeline to over 20,000 Android apps sourced from multiple app markets, we detected 2,357 apps exhibiting device-specific behaviors.
In terms of distribution, we found that apps from Chinese app markets in China exhibit more device-specific behaviors compared to apps on Google Play, and device-specific behaviors are more likely to involve manufacturers with a larger market share.
In terms of categorization, we have classified these device-specific behaviors into 29 types according to their functionalities.
These device-specific behaviors are predominantly implemented to fix compatibility issues or adapt to manufacturer-introduced new features.
Additionally, we discovered privacy-related device-specific behaviors, revealing that 33 aggressive apps (including popular ones with millions of downloads) are abusing the customization permission mechanism to obtain user-unresettable identifiers without requiring any permissions, which should be protected by system-level permissions, posing a substantial threat to user privacy.
We further investigated the sources of these device-specific behaviors, and the channels through which developers can gather information related to device-specific behaviors.
Notably, we encountered a scarcity of relevant information in the official documentation provided by manufacturers.
Through our research, we heighten developers' awareness of the extensive range of device-specific behaviors and provide guidance on code practices for implementing such behaviors. Additionally, our study highlights areas where manufacturers could improve their official documentation regarding device-specific behaviors and exercise greater caution to prevent the introduction of new security risks during system customization.

In summary, this work offers these major contributions:

\begin{itemize}
    \item We take the first step to present a comprehensive analysis of device-specific behaviors in large-scale real-world Android apps by implementing an automated static analysis framework, and we have successfully uncovered 2,357 apps with such behaviors.
    \item 
    We make effort to characterize these device-specific behaviors, and create a taxonomy of 29 categories. Beyond the common practices including issue fix and feature adaptation, we have identified some aggressive behaviors that breach AOSP protection to obtain sensitive information unscrupulously, including some popular apps.
    \item
    Our research empowers developers with a more profound insight into device-specific behaviors, while underscoring the imperative for manufacturers to enhance relevant official documentation and fortify the security of system customization.
\end{itemize}

We make the source code of our tool, the summarized rules, and the dataset publicly available~\cite{opensource} to the research community.

\section{Background}
\label{sec:background}

\subsection{Fragmentation Issues}
Android fragmentation, arising from device variability and manufacturer customizations, poses significant challenges. 
Manufacturers leverage Android's open-source nature to customize system versions from AOSP, introducing new features like AI assistants~\cite{oppoai} and mini window functionality~\cite{vivominiwindow} to differentiate their devices. 
These alterations and customizations amplify distinctions among various devices, furthering the phenomenon of device-specific fragmentation.

While fragmentation offers choice to consumers, it complicates app development, as app behavior varies across devices, especially with customized UI (e.g., home screen and system bar) and hardware components (e.g., camera and Bluetooth).
Ensuring seamless performance with various system and hardware configurations across a wide spectrum of Android devices, and harnessing the novel features in customized systems, demands developers to dedicate considerable time and financial resources to the development and testing process.

\subsection{Privacy-related System Customizations}
A unique class of system customizations pertains to privacy-related modifications. 
As certain manufacturers extensively customize the AOSP system, these customizations can extend to components associated with user privacy, including components linked to device identifiers or user location information. 
When issues emerge in such customizations, they introduce additional security risks to the customized system, consequently jeopardizing user privacy.
Manufacturers might exhibit delays in adopting Google's measures aimed at safeguarding user privacy~\cite{el2021dissecting}. 
An illustrative instance is observed in Android 10, where Google restricts third-party apps from accessing non-resettable identifiers like the IMEI. 
Nevertheless, LG's system did not promptly revise the access control policy for custom IMEI access APIs, resulting in a scenario where apps can still obtain the IMEI of LG devices even under the Android 10 system.
Moreover, insufficient security measures may result in the exposure of user privacy data through newly introduced features by manufacturers~\cite{hou2022large}. 
In the case of the customized weather app by Vivo, the absence of security protection permits the app to access the user's accurate location without the necessity of seeking any permissions.
Unscrupulous developers may exploit vulnerabilities in these customized systems through reverse engineering, potentially compromising user privacy data that ought to be safeguarded.

\section{Approach}
\label{sec:tool}

\begin{figure*}[ht]
  \centering
  \resizebox{0.8\textwidth}{!}{
  \includegraphics[width=\linewidth]{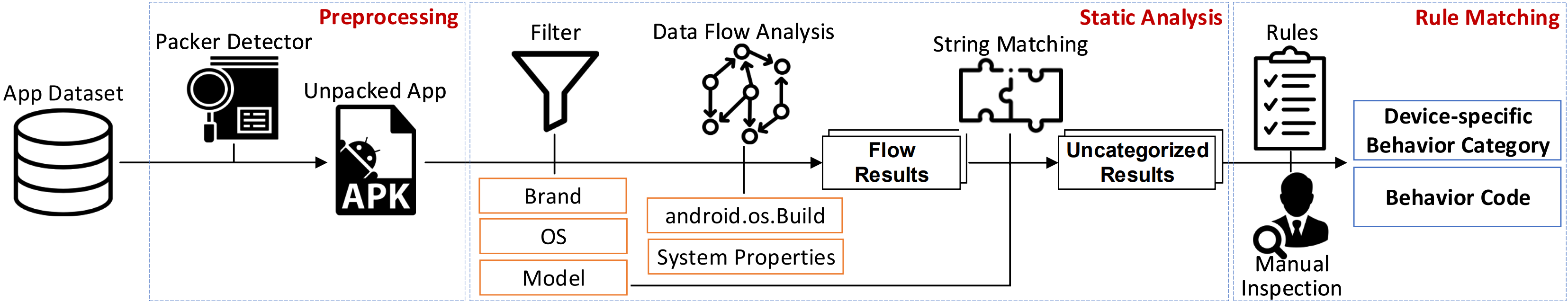}
  }
  \caption{Study Methodology Design.}
  \label{fig:approach}
  \vspace{-0.1in}
\end{figure*}

To gain a holistic understanding of device-specific behaviors in actual Android apps, we meticulously designed research approaches, as illustrated in \autoref{fig:approach}.
The process takes an APK (i.e., Android Package) as input and outputs the app's device-specific category (e.g., user interface and permission management) along with the corresponding behavior code.
Our approach contains three key steps:
(1) \textit{Preprocessing}, which detects whether apps employ packing techniques, and selects unpacked apps as analysis targets; 
(2) \textit{Static Analysis}, which utilizes data flow analysis techniques to identify code snippets related to device-specific behaviors; 
(3) \textit{Rule Matching}, which involves categorizing the result from the second step using rules obtained through manual inspection and generating an analysis result, including behavior category and relevant code.

\subsection{Preprocessing}

In order to enhance the security of apps, some developers may employ app packing techniques, which involve anti-static analysis (e.g., encryption, obfuscation and virtual machine-based protection) and anti-dynamic analysis (e.g., running environment detection, anti-debugging and seizing debugging interface) and anti-tampering (e.g., file integrity verification)~\cite{dong2022did, xue2020packergrind, duan2018things}. 

App packing techniques can effectively prevent program analysis and cracking all existing packing solutions remains a formidable challenge. 
Consequently, our analysis concentrates solely on unpacked apps, and packed ones are filtered out. 
Employing established packing detection tools~\cite{apkprotectionsearch}, we verify the packing status of apps. This process is based on scrutinizing featured files or package names within the APK files.

\subsection{Static Analysis}
We utilize static analysis techniques to scrutinize device-specific behaviors in apps. 
This approach allows us to conduct a comprehensive examination of device-specific behaviors within apps without the necessity of preparing numerous devices from diverse brands and models.

\noindent \textbf{Observations based on pilot study.}
Our objective is to detect all methods that may encompass device-specific behaviors. 
To accomplish this, we initially outline the pattern of device-specific behaviors. 
Following a reverse analysis and examination of over 100 real-world apps, we have deduced the following observations:

\begin{table}[ht]
\caption{The Method to Obtain Device Information.}
\label{table:source}
\resizebox{1\linewidth}{!}{
\begin{tabular}{|l|l|l|}
\hline
\textbf{Field in android.os.Build} & \textbf{System Property}       & \textbf{Description}      \\ \hline
BRAND                     & ro.product.brand        & Consumer-visible Brand                           \\ \hline
DEVICE                    & ro.product.device       & Name of the Industrial Design                    \\ \hline
DISPLAY                   & ro.build.display.id     & Build ID String for Users                        \\ \hline
FINGERPRINT               & ro.build.fingerprint    & String that Identifies Current Build             \\ \hline
MANUFACTURER              & ro.product.manufacturer & Manufacturer of the Product/Hardware             \\ \hline
MODEL                     & ro.product.model        & End-user-visible Name for the Product            \\ \hline
PRODUCT                   & ro.product.name         & Name of the Overall Product                      \\ \hline
\end{tabular}
}
\vspace{-0.1in}
\end{table}

(a) Prior to engaging in device-specific behaviors, the app must initially retrieve current device information, such as the brand and specific device model. 
As elucidated in the motivation section, there are two approaches for obtaining this device information: one entails accessing the fields within the \texttt{\seqsplit{android.os.Build}} class,  which contains information derived from system properties about the current environment. 
The alternative method involves directly retrieving the values of system properties associated with the device information, typically achieved by invoking methods in the \texttt{\seqsplit{android.os.SystemProperties}} class through Java reflection.
We list the relevant methods in \autoref{table:source}.
(b) After acquiring the current device information, the app assesses whether the current environment necessitates the execution of device-specific behaviors. 
Consequently, the code related to device-specific behaviors contains identifiers (typically in string form) for devices requiring special handling by the app, including specific brands, operating systems, and models. 
These identifiers are subsequently employed in comparisons with the current device information.
(c) When comparing the current device information with target device identifiers, apps often employ “if statements” and string processing functions, such as functions \texttt{\seqsplit{equal}}, \texttt{\seqsplit{toLowerCase/toUpperCase}}, and \texttt{\seqsplit{startsWith/endsWith}}. 
To conclude, the device-specific behavior code should include “if statements”, with the original or processed device information incorporated into the conditions of these “if statements”.

\vspace{0.1em}
\noindent \textbf{Methodological steps in detail.}
Based on our observations, we formulate the following static analysis process.

First, we constructed a comprehensive database containing various device information that is globally available. For this purpose, we crawled GSMArena~\cite{gsmarena}, a popular online platform dedicated to mobile technology and smartphones, which includes detailed information about devices from different manufacturers.
Eventually, the information we collected includes 72 brand names (e.g., Samsung, Xiaomi, and Huawei), 33 operating system names (e.g., One UI, MIUI, and HarmonyOS), and 8,032 mobile model names (e.g., SM-S918B represents Samsung Galaxy S23 Ultra, and ALN-AL00 refers to Huawei Mate 60).

\begin{figure}[h]
  \centering
  \vspace{-0.15in}
  \resizebox{0.8\linewidth}{!}{
  \includegraphics[width=\linewidth]{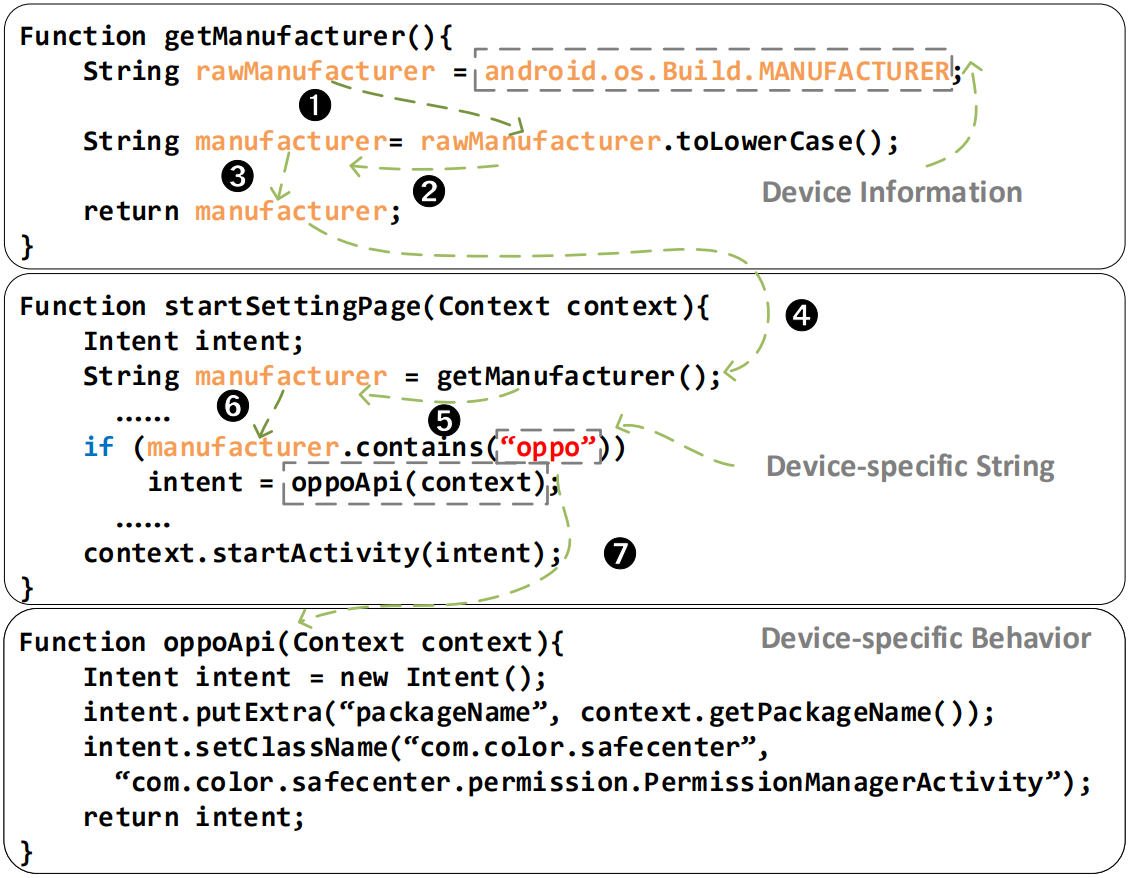}
  }
  \caption{Inter-procedure Data Flow Analysis.}
  \label{fig:dataflow}
  \vspace{-0.1in}
\end{figure}

Then, we aim to conduct inter-procedural data flow analysis.
We decompile APK files and transform Java code into \texttt{\seqsplit{Jimple}}, which is a three-address intermediate representation, and generate Inter-process Control Flow Graph (ICFG) for apps to obtain the function call relationships.
Our approach is implemented on top of Soot~\cite{vallee2010soot} and FlowDroid~\cite{arzt2014flowdroid}, both of which are widely used Android static analysis tools.
We manage to identify all methods capable of accessing current device information according to our first observation.
To be more specific, we start with traversing all methods in the app to locate the two methods (i.e., fields in \texttt{\seqsplit{android.os.Build}} class and system properties). After identifying a method that acquires device information, 
we conduct an intra-procedure def-use analysis on variables storing device information to find all uses of the variable within the method, as shown in Steps~\ding{202} and \ding{204} in \autoref{fig:dataflow}.
The example in \autoref{fig:dataflow} illustrates the process of starting the permission settings page. 
Firstly, the startSettingPage function obtains the current device manufacturer through the getManufacturer function. 
Upon determining that the current device brand is OPPO, it invokes the \texttt{oppoApi} function to navigate to OPPO's permission settings page.

Further, we proceed with inter-procedural analysis when encountering the following scenarios:
(a) When device information is passed as a parameter to a function call, we employ different strategies for inter-procedural data flow analysis depending on the type of calling function. 
In instances where the called methods have concrete bodies, we perform intra-procedural data flow analysis on the corresponding parameters in the called method. 
In cases where the called methods lack concrete method bodies, such as Java library methods (e.g., string processing methods), we skip the analysis of the method body. 
Instead, we directly perform intra-procedural data flow analysis on the method's return value within the original method, as illustrated by Step~\ding{203} in \autoref{fig:dataflow}.
(b) When device information, including its variants, is used as a method's return value, such as Step~\ding{204} in our example. 
We utilize the ICFG to identify all methods calling the method that return device information, which is depicted by Step~\ding{205}.
Subsequently, we perform intra-procedural data flow analysis on the variables corresponding to the method's return values within these calling methods, as shown in Steps~\ding{206} and \ding{207}.

Having obtained all the methods capable of accessing device information, We aim to identify ``if statements'' used to compare the current device information with the target device identifiers based on our second and third observations.
With the results of the aforementioned data flow analysis, we can determine whether the current device information appears within the condition of an ``if statement'', as illustrated in Step~\ding{207} in \autoref{fig:dataflow}.
Next, we check whether the target device identifier is present around the founded if statement using string matching and our device information database. \looseness=-1

After confirming that a specific if statement compares the current device information with the target device identifier, we conclude that the branches controlled by the if statement contain device-specific behaviors.
If there are method invokes within the branch, and the called methods have concrete bodies, we further traverse the called methods and consider them as device-specific behaviors, as shown in Step~\ding{208} in \autoref{fig:dataflow}.
When new method calls still can be found during this process, we continue the deepening process until no new method calls are discovered.
Consequently, we identify all code snippets related to device-specific behaviors.

\begin{table*}[thb]
\caption{Examples of the 29 categories of the identified rules.}
\label{table:rules}
\resizebox{1\linewidth}{!}{
\begin{tabular}{|c|c|l|c|c|l|c|c|l|}
\hline
\textbf{Behavior Type}                                                                       & \textbf{Category}                                                                          & \textbf{Keyword}                                                            & \textbf{Behavior Type}            & \textbf{Category}                                                                         & \textbf{Keyword}    & \textbf{Behavior Type}                    & \textbf{Category}     & \textbf{Keyword}    \\ \hline
\multirow{4}{*}{\begin{tabular}[c]{@{}c@{}}Compatibility \\ Issue Fix\end{tabular}} & \multirow{2}{*}{\begin{tabular}[c]{@{}c@{}}Memory\\ Leak\end{tabular}}            & \begin{tabular}[c]{@{}l@{}}android.view.inputmethod.\\ InputMethodManager\end{tabular} & \multirow{4}{*}{\begin{tabular}[c]{@{}c@{}}Manufacturer-introduced\\ Feature Adaptation\end{tabular}} & \multirow{2}{*}{\begin{tabular}[c]{@{}c@{}}Permission\\ Management\end{tabular}} & com.vivo.permissionmanager & \multirow{4}{*}{Privacy-related} & \multirow{2}{*}{OAID} & com.huawei.hwid        \\ \cline{3-3} \cline{6-6} \cline{9-9}  &            & \begin{tabular}[c]{@{}l@{}}android.gestureboost.\\ GestureBoostManager\end{tabular}    &   &   & com.miui.securitycenter    &            &              & com.asus.msa           \\ \cline{2-3} \cline{5-6} \cline{8-9} & \multirow{2}{*}{\begin{tabular}[c]{@{}c@{}}Malfunctional\\ Features\end{tabular}} & android.webkit.WebSettings        &      & Foldables    & hardware.sensor.posture    &          & \multirow{2}{*}{\begin{tabular}[c]{@{}c@{}}SystemProperties\\ Containing Hardware Identifiers\end{tabular}} & ro.meizu.hardware.imei \\ \cline{3-3} \cline{5-6} \cline{9-9}   &           & android.bluetooth.BluetoothAdapter         &          & DisplayCutouts   & flyme.config.FlymeFeature  & & & ro.ril.miui.meid       \\ \hline
\end{tabular}
}
\vspace{-0.1in}
\end{table*}

\subsection{Rule Matching}
\label{sec:rulematching}
After isolating the code snippets related to device-specific behaviors, we adopt a rule-matching methodology for categorizing these behaviors based on the functionality implemented by the code. After the static analysis phase, we encountered an extensive amount of code snippets associated with device-specific behaviors, making it challenging for individual inspection of each snippet. Nevertheless, we observe that certain device-specific behaviors, such as addressing prevalent issues or adapting to common features, frequently recur across numerous apps. In light of this, we have implemented a clustering method to facilitate our categorization.

We cluster code snippets that invoke identical system methods.
Such a decision is based on our understanding that (a) apps must utilize these system methods to implement corresponding system functionalities, (b) code snippets that call the same system methods share similar functionalities, and (c) calls to these system methods would remain unchanged after code obfuscation in general.
After the clustering, we sample code snippets from each cluster, attempting to manually select keywords that best differentiate the functionality of this cluster to formulate our rule.
These keywords are generally sourced from method names, strings containing method names (commonly used in intent construction or Java reflection), or other meaningful strings (e.g., customized system property names), considering the rich semantic information they provide.
For example, in the case shown in~\autoref{fig:dataflow}, the string \texttt{\seqsplit{com.color.safecenter.permission.PermissionManagerActivity}} is chosen as the keyword, which is the activity name corresponding to the OPPO permission settings page.
Each selected keyword is then manually labeled with a functionality category (e.g., permission) to formulate the rules in the form of a key-value pair, i.e., {\texttt{[Category:Keyword]}}.
If the code snippet contains the relevant keyword, we categorize it into the corresponding category.

The rule-labeling process was primarily led by two experienced Android researchers, with a third for dispute resolution. Specifically, the two lead researchers independently reviewed code snippets and formulated rules. They then convened to discuss and reconcile their findings. Rules that were mutually agreed upon were adopted. In instances of discord, the third researcher was consulted to resolve the conflict. 

In total, we labeled 170 keyword rules corresponding to 29 distinct categories.
Naturally, one functionality category may correspond to multiple keywords.
We list some of the rule examples in \autoref{table:rules}, with the detailed breakdown available on our website~\cite{opensource}.
These categories are aligned with the three major behavior types outlined in the Introduction, i.e., compatibility issue fix (e.g., memory leak and malfunctional features), manufacturer-introduced feature adaptation (e.g., permission management and UI-related adaptations), and privacy-related device-specific behaviors (e.g., system properties leaking hardware identifiers). 
We elaborate them in \S\ref{sec: RQ2}.

\section{Results}
\label{sec:result}

\subsection{Dataset}
To explore device-specific behaviors in large-scale real-world Android apps, we amassed a total of 21,889 different apps from Google Play and two third-party app markets, namely App China~\cite{appchina} and Huawei AppGallery~\cite{appgallery}.
Given Google Play's dominance in the global market, we first sourced Android apps directly from it, utilizing the widely recognized AndroZoo~\cite{allix2016androzoo} dataset.
Our focus was on selecting the most recently updated apps from this extensive collection.
Another driving factor was the unavailability of Google services in China, coupled with the significant presence of prominent device manufacturers from China. 
This prompted us to conduct a comprehensive global study on device-specific behaviors, leading us to include App China. 
App China, a dedicated Android app market, is focused on serving users in China with a substantial number of apps available in AndroZoo~\cite{appchinaandrozoo}.
Furthermore, to explore potential brand biases in manufacturer-organized app markets, where apps may receive more attention from manufacturer systems, we selected the Huawei AppGallery as a representative case for investigation.
We employed crawlers to gather apps from two third-party markets, and the details of app distribution and device-specific behaviors are shown in \autoref{table:marketsplit}.

\begin{table}[h]
\vspace{-0.1in}
\caption{Distribution of Apps Collected and Unpacked Apps Across Markets}
\label{table:marketsplit}
\resizebox{1\linewidth}{!}{
\begin{tabular}{|c|r|r|r|}
\hline
                  & \textbf{\# Apps Collected} & \textbf{\# Unpacked Apps} & \textbf{\# Apps with Behaviors} \\ \hline
Google Play       & 12,469            & 12,255           & 1,147                 \\ \hline
App China         & 3,120             & 1,758            & 443                   \\ \hline
Huawei AppGallery & 6,300             & 2,126            & 767                   \\ \hline
Total             & 21,889            & 16,139           & 2,357                 \\ \hline
\end{tabular}
}
\vspace{-0.1in}
\end{table}

\subsection{Research Questions}
In our exploratory investigation aimed at comprehending the current state of device-specific behaviors in real-world Android apps, we strive to address the following three research questions (RQs):

\begin{itemize}
    \item[RQ1] What is the distribution of device-specific behaviors in real-world Android apps?
    Does the prevalence of such behavior vary significantly across different app markets?
    What is the distribution of the functionalities associated with these behaviors?
    \item[RQ2] What specific functionalities do these device-specific behaviors implement? 
    \item[RQ3] 
    Where did these device-specific behaviors come from? From the developers' standpoint, how can they acquire knowledge about these device-specific behaviors? 
\end{itemize}

\subsection{RQ1: Distribution of Device-specific Behaviors}
\label{sec:rq1}
After experimenting with 16,139 unpacked real-world Android apps, we successfully analyzed 7,079 of them, discovering device-specific behaviors in 2,357 of the apps.
Our experiment was unable to analyze over half of the collected apps, primarily due to our testing targets being real-world apps from app markets, which is relatively complex, and limitations of the static analysis tools used in our experiment, as mentioned in previous studies~\cite{bleier2023ahead, zhang2021analyzing}.
In~\cite{zhang2021analyzing}, when analyzing apps from Google Play using FlowDroid, the analysis was successful for 56\% of the apps, yielding results similar to ours.
The static analysis tool encounters specific step analysis failures. 
Upon the occurrence of this exception, the tool continuously retries the corresponding steps, entering an infinite loop and ultimately reaching the 1-hour timeout we have set.
In terms of the proportion of apps exhibiting device-specific behaviors, the two app markets from China (i.e., 70\% in Huawei AppGallery and 77\% in App China) are significantly higher than that of Google Play (i.e., 21\% in Google Play).
From this, we can conclude that developers from China tend to incorporate more device-specific behaviors into their apps.

\begin{figure}[h]
  \centering
  \vspace{-0.1in}
  \resizebox{0.8\linewidth}{!}{
  \includegraphics[width=0.9\linewidth]{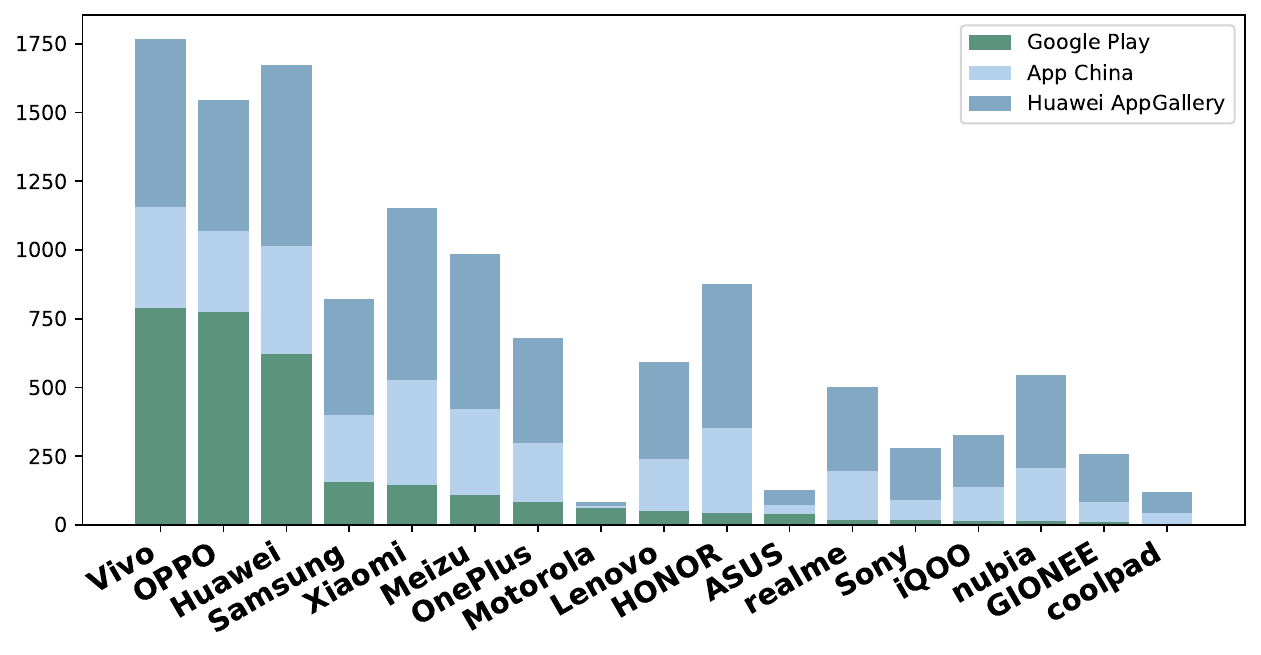}
  }
  \begin{tablenotes}
	\footnotesize
	\item Note: from left to right in the figure, the device-specific behaviors are sorted by the frequency of occurrence in Google Play apps. This facilitates a comparison of the differences between Google Play and the other two app markets. The sorting method used in \autoref{fig:os} and \autoref{fig:matching} is the same as in this figure.
  \end{tablenotes}
  \caption{The Distribution of Brands Involved in Device-specific Behaviors.}
  \label{fig:brand}
\end{figure}

\begin{figure}[h]
  \centering
  \resizebox{0.7\linewidth}{!}{
  \includegraphics[width=\linewidth]{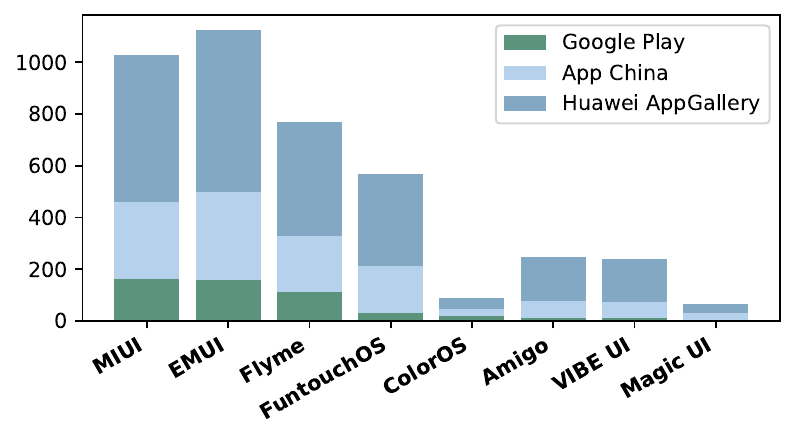}
  }
  \vspace{-0.1in}
  \caption{The Distribution of OSes Involved in Device-specific Behaviors.}
  \label{fig:os}
  \vspace{-0.1in}
\end{figure}

\begin{figure}[h]
  \centering
  \resizebox{1\linewidth}{!}{
  \includegraphics[width=\linewidth]{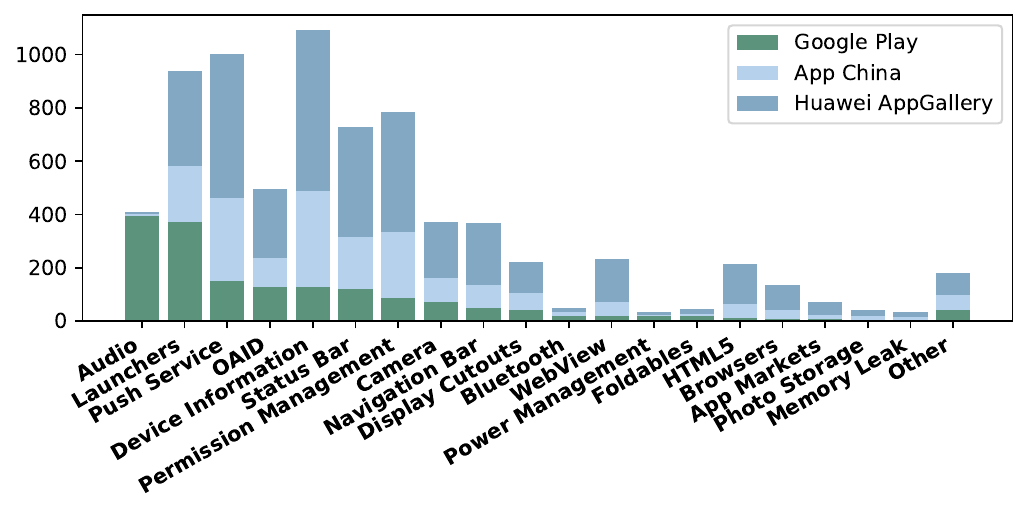}
  }
  \vspace{-0.3in}
  \caption{The Distribution of Functionalities Implemented by Device-specific Behaviors.}
  \vspace{-0.1in}
  \label{fig:matching}
\end{figure}

Furthermore, in \autoref{fig:brand} and \autoref{fig:os}, we summarized the distribution of brands and operating systems involved in these device-specific behaviors.
In general, brands with high market share and their corresponding operating systems are more likely to appear in device-specific behaviors.
Apps from two Chinese markets tend to implement device-specific behaviors for a broader range of brands and customized systems.
The notable observation is that the prevalence of Vivo, OPPO, and Huawei devices in Google Play apps surpasses that of other brands significantly.
This is because many Google Play apps tested leverage the customized audio features in the systems of these three manufacturers when developing audio-related functionalities. 
Further, we did not observe such behavior in apps from the two Chinese markets, as illustrated in the first column of \autoref{fig:matching}.
Moreover, through a comparison of the results from App China and the Huawei AppGallery, we did not observe significant brand bias.
It's worth noting that we have excluded brands and operating systems that appeared less than 20 times in \autoref{fig:brand} and \autoref{fig:os}.
We did not observe many device-specific behaviors related to Huawei's HarmonyOS because
identifying the HarmonyOS system requires dynamically invoking custom APIs within the HarmonyOS system through Java reflection, which is beyond the capability of our approach.

In \autoref{fig:matching}, we have statistically analyzed the distribution of device-specific behaviors based on their functionalities.
After a manual inspection in \S\ref{sec:rulematching}, we formulated a total of 29 rules, with each rule corresponding to a specific behavioral feature.
The most prevalent device-specific behavior in Google Play apps involves the previously mentioned audio features, which, however, appear infrequently in the other two third-party app markets.
In terms of other functionalities, the differences between Google Play and other app markets are relatively smaller. 
Apps from the two third-party app markets exhibit more instances of device-specific behaviors.
Additionally, we also calculated the average number of brands, systems, and models involved in apps with device-specific behaviors in each app store, as shown in \autoref{table:average}.
In the Chinese third-party app markets, a single app tends to include more device-specific behaviors, involving a greater variety of brands, systems, and models, and implementing a wider range of functionalities. 
This explains why, despite having fewer apps with device-specific behaviors in the two third-party markets compared to Google Play, they exhibit a higher diversity of device-specific behaviors.

\begin{table}[h]
\vspace{-0.2in}
\caption{The Average Number of Involved Device Information and Functionalities in Apps with Device-specific Behaviors in Each App Market.}
\label{table:average}
\resizebox{1\linewidth}{!}{
\begin{tabular}{|c|r|r|r|r|}
\hline
                  & \textbf{Avg Brands} & \textbf{Avg OSes} & \textbf{Avg Models} & \textbf{Avg Functionalities} \\ \hline
Google Play       & 2.904      & 0.493    & 1.150      & 2.267               \\ \hline
App China         & 7.365      & 4.130    & 2.541      & 9.939               \\ \hline
Huawei AppGallery & 7.522      & 4.577    & 3.027      & 10.778              \\ \hline
\end{tabular}
}
\end{table}
\vspace{-0.2in}

~\\
\noindent \fbox{
\parbox{0.45\textwidth}{
\textbf{Answer to RQ1:} 
\textit{Device-specific behaviors are prevalent in real-world apps, especially in the two third-party app markets, where their prevalence surpasses that of Google Play. 
Mainstream brands exhibit these behaviors more frequently, and manufacturer-organized app markets show no brand bias.}
}}

\subsection{RQ2: Functionality Implemented through Device-specific Behaviors}
\label{sec: RQ2}

Device-specific behavior can be broadly categorized into three major types, including 29 kinds of sub-categories.
\subsubsection{Compatibility Issue Fix.}
Due to manufacturers' customization of device hardware and system components, certain devices may experience specific issues.
These issues are usually discovered in a particular system version and addressed in one or several subsequent system updates.
While these issues are typically resolved within a relatively short period, the method of system updates does not guarantee that the solutions will reach every user's device in a timely manner.
Not all users undertake system updates promptly, leading to a significant extension of the lifecycle of these issues. This necessitates app developers to address these device-specific issues.
Device-specific issues can be further classified into two types: memory leak issues and malfunctional features.

\vspace{0.1em}
\noindent\textbf{Memory Leak.}
We have discovered that some device-specific behaviors aim to address memory leak issues in specific components of Huawei (i.e., EMUI) and Samsung systems in certain versions.
Memory leak is a type of resource leak that occurs when a program does not timely release unneeded memory, leading to a gradual reduction in available system memory.
In Android 7 EMUI 5, there is a memory leak issue in the EMUI system's customized component \texttt{\seqsplit{android.gestureboost.GestureBoostManager}}.
This component holds a reference to the app's activity and fails to release it when the app stops and garbage collection is needed, leading to memory leaks.
Similarly, Samsung's system components \texttt{\seqsplit{com.samsung.android. content.clipboard.SemClipboardManager}} and \texttt{\seqsplit{com.samsung.android.emergencymode.SemEmergencyManager}} in Android 7, along with Huawei's \texttt{\seqsplit{android.view.inputmethod.InputMethodManager}} in Android 8, all contain memory leak issues.
Developers need to manually release references from the problematic system components to the app components (e.g., Activities) when app components are destroyed, allowing the garbage collection to proceed correctly.

\vspace{0.1em}
\noindent \textbf{Malfunctional Features}. 
Another type of device-specific issue is that certain system features fail to function properly on some devices.
The system features refer to those present in the AOSP system, so this exceptional situation will have an impact on developers' normal development. 
We further categorize them based on features into the following categories:

\textbf{(1) Camera.}
The camera, as a crucial component of smartphones, receives significant attention from manufacturers, who invest considerable efforts in customizing camera hardware.
But this has also led to many malfunctioning features.
For example, Lenovo devices with the model X804F cannot reliably retrieve the ID of the front camera through the \texttt{\seqsplit{android.hardware.Camera}} class.
The Nexus 5X and Mi 5 have the opposite rotation direction when setting the orientation of photo previews compared to other devices.
When setting the focus mode for the camera, some devices (i.e., MEIZU MX3 and Samsung Galaxy S4) cannot retrieve their supported focus modes through system APIs and must rely on developers to set the focus modes individually.

\textbf{(2) Photo Storage.}
In the AOSP system and on most devices, camera photos are stored in the system path ``/sdcard/DCIM/Camera'', while screenshots are stored in ``/sdcard/Pictures/Screenshots''.
But Xiaomi has changed the storage path of screenshots to ``/sdcard/DCIM/Screenshots'', requiring developers to make separate modifications for Xiaomi devices when obtaining screenshot files through file paths.

\textbf{(3) HTML5.}
The issues related to HTML5 mainly revolve around the HTML5 geolocation feature, which enables apps to obtain the device's geographical location through JavaScript code.
In some devices running a specific version of the Android, the HTML Geolocation feature is not available, and this seems not to be determinable through system APIs. 
Therefore, some developers have compiled a list of devices, along with the corresponding system versions, (e.g., Vivo Xplay5A running Funtouch OS\_2.5.1) 
where this feature cannot be used.

In other aspects, the overview mode of \textbf{WebView} on Meizu devices is different from that on other devices and does not function properly.
In terms of \textbf{Bluetooth} functionality, some devices require special parameters during Bluetooth device scanning, and there are issues with the offloaded scan batching and offloaded filter functionality on certain devices.
The \textbf{VPN} functionality on Samsung devices may ignore certain DNS servers, requiring apps to manually add routes for DNS servers.

\subsubsection{Manufacturer-introduced Feature Adaptation.}
Device manufacturers, building upon Google's AOSP system, add new features to their customized systems to enhance product competitiveness. 
Faced with these manufacturer-introduced features, app developers need to undertake substantial adaptation efforts to fully leverage these features, thereby implementing app functionality and enhancing user experience.

\vspace{0.1em}
\noindent\textbf{Push Service.}
Push services serve as mechanisms that facilitate apps in delivering message notifications to user devices, even when the device is locked, and the app is not actively running. 
Google offers the official push notification service, Firebase Cloud Messaging (FCM)~\cite{fcm}, which is a widely adopted solution. 
However, in certain countries like China, where devices cannot connect to Google services, utilizing FCM for push notifications becomes unfeasible.
Developers have the option to implement message push through long connections, where a lasting link is upheld between the device and the server. 
Furthermore, device manufacturers introduce their proprietary push services, such as Huawei Push Kit~\cite{huaweipush} and Mi Push~\cite{mipush}. 
Developers can integrate the manufacturer's push SDK into their app, enabling the creation of push notifications on the manufacturer's operating platform, which can also provide enhanced stability.
Hence, we observed that developers incorporate push SDKs from prominent device manufacturers into their apps, employing the respective manufacturer's push service depending on the current system.

\vspace{0.1em}
\noindent\textbf{Power Management and Permission Management.}
In today's smartphone realm, battery life is paramount in assessing device performance. 
Manufacturers enhance battery capacity while optimizing systems to reduce unnecessary code execution.
Yet, stringent power management policies pose challenges for developers. 
In stock Android, apps may persist in the background despite user closure via the recent screen~\cite{recentscreen}, initiating actions like receiving push notifications.
Background app operation ceases only with force stop or manual battery optimization settings.
For example, apps using Firebase for push won't receive notifications under these conditions.
Conversely, customized systems force-stop apps upon recent screen closure, disallowing background operation by default. 
Hence, developers request auto-start permissions to ensure functionality like push messaging, absent in stock Android. 
Consequently, managing permissions lacks an official method, requiring apps to navigate specific activities to guide users, as depicted in \autoref{tab:autostart}. 
Similarly, some manufacturers revamp permission settings pages (i.e., different activity) to handle custom permissions and provide a visually appealing user interface.

\begin{table}[h]
\vspace{-0.1in}
\caption{The Activities for Managing App Auto-start Permissions.}
\label{tab:autostart}
\resizebox{1\linewidth}{!}{
\begin{tabular}{|l|l|}
\hline
\textbf{Manufacturer} & \textbf{Activity for Managing Auto-start   Permissions}                      \\ \hline
Samsung      & com.samsung.android.sm.ui.battery.BatteryActivity                   \\ \hline
Xiaomi       & com.miui.permcenter.autostart.AutoStartManagementActivity           \\ \hline
ASUS         & com.asus.mobilemanager.powersaver.PowerSaverSettings                \\ \hline
Huawei       & com.huawei.systemmanager.startupmgr.ui.StartupNormalAppListActivity \\ \hline
OPPO         & com.coloros.safecenter.permission.startup.StartupAppListActivity    \\ \hline
vivo         & com.vivo.permissionmanager.activity.BgStartUpManagerActivity        \\ \hline
Nokia        & com.evenwell.powersaving.g3.exception.PowerSaverExceptionActivity   \\ \hline
OnePlus      & com.oneplus.security.chainlaunch.view.ChainLaunchAppListActivity    \\ \hline
\end{tabular}
}
\vspace{-0.1in}
\end{table}

\noindent\textbf{Launchers.}
Badge refers to a visual indicator displayed on the app icon on the home screen (i.e., launchers).
It is often a small numerical count or a dot, providing a glance at the status of the app, such as the number of unread messages or notifications.
However, the stock Android system does not support displaying numbers on the badge.
Influenced by iOS, which supports displaying numbers in the badge, major device manufacturers have implemented the functionality to show numbers in the badge within their customized launchers.
Developers need to call the corresponding customized APIs to modify the number in the badge.

Another issue related to the launchers is shortcuts~\cite{shortcuts}.
Different customized systems have varying approaches to shortcut management. 
Google and some Samsung devices, by default, allow apps to create shortcuts, while certain manufacturers (such as Huawei and Meizu) do not grant this permission. 
Apps need to adopt different methods based on the system to check the permission status for install shortcuts.

\vspace{0.1em}
\noindent\textbf{Display Cutouts and Foldables.}
A display cutout or notch refers to an area on some devices that extends into the display surface. 
Although manufacturers like Huawei, Xiaomi, and OPPO had already introduced devices with cutouts upon the release of Android 8, official support for them via the \texttt{\seqsplit{DisplayCutout}} class was included in Android 9.
Initially, these manufacturers incorporated cutout adaptation into their customized Android 8 systems, providing position and dimensions for the cutouts. Subsequently, with the launch of Android 9, they shifted to the official adaptation method. 
However, since many users delay system updates, developers must implement diverse strategies based on the Android version to ensure proper cutout support. They rely on manufacturers' methods for Android 8 and adopt the official approach for versions post-Android 9.
Another emerging design trend is foldable~\cite{fordable}.
For effective utilization, apps need to adjust layouts based on foldable states (e.g., flat and half-opened). Manufacturers store this information in various ways (e.g., custom APIs and system properties).

\vspace{0.1em}
\noindent\textbf{Device Information.}
When manufacturers customize based on the AOSP system, similar to AOSP, they release multiple sub-versions within a major version. 
Manufacturers may address some issues in minor version updates. 
To access information about these issue fixes or simply to collect device information for analytics, developers need to obtain the version of the customized system.
The version information is typically stored in customized system properties. 

Beyond the aforementioned considerations, a parallel situation exists with the \textbf{Status Bar}, mirroring the challenges posed by display cutouts. 
Manufacturers introduced new styles for the status bar prior to Android, necessitating developers to employ diverse methods for distinct system versions. 
Conversely, certain manufacturers introduce novel \textbf{Navigation Bar} styles, including Meizu's smart bar. 
Apps must utilize customized functions to retrieve the status of the system bar and adapt the interface display accordingly.
The manufacturers also add some new pre-installed apps, such as \textbf{Browsers}, \textbf{Video Players}, and \textbf{App Markets}. 
When needed, apps can leverage these more powerful customized apps.
As mentioned in \S\ref{sec:rq1},  some manufacturers provide additional \textbf{Audio} features, offering apps with ear return functions.

\subsubsection{Privacy-related.}
\label{sec:privacyresult}
Furthermore, certain device-specific behaviors are employed to gather sensitive user data, raising potential privacy concerns.

\vspace{0.1em}
\noindent\textbf{OAID.}
Starting with Android 10, third-party apps are restricted from accessing non-resettable hardware identifiers like IMEI, IMSI, and serial numbers. 
As an alternative, Google recommends apps use advertising identifiers~\cite{adid} for advertising and analytics purposes.
However, in regions where access to Google services is restricted, Google's advertising identifier also becomes unfeasible. 
To address this gap, in 2019, the China Mobile Security Alliance introduced the OAID (Open Anonymous Device Identifier)~\cite{oaid} as an alternative.
Unlike Google's advertising identifier, OAID's implementation varies by manufacturer and is embedded in their customized systems. 
Consequently, methods for obtaining OAID differ across devices, necessitating developers to adapt to these discrepancies.

For instance, Samsung utilizes the \texttt{\seqsplit{com.samsung.android.deviceidservice.DeviceIdService}}, Xiaomi employs the API \texttt{\seqsplit{com.android.id.impl.IdProviderImpl.getOAID}}, and Huawei uses the system setting \texttt{\seqsplit{pps\_oaid}}.
As each manufacturer independently implements OAID, there's a risk of vulnerabilities or inadequate security measures, potentially compromising user privacy~\cite{blackhat}.

\begin{figure}[h]
  \centering
  \vspace{-0.1in}
  \resizebox{0.9\linewidth}{!}{
  \includegraphics[width=\linewidth]{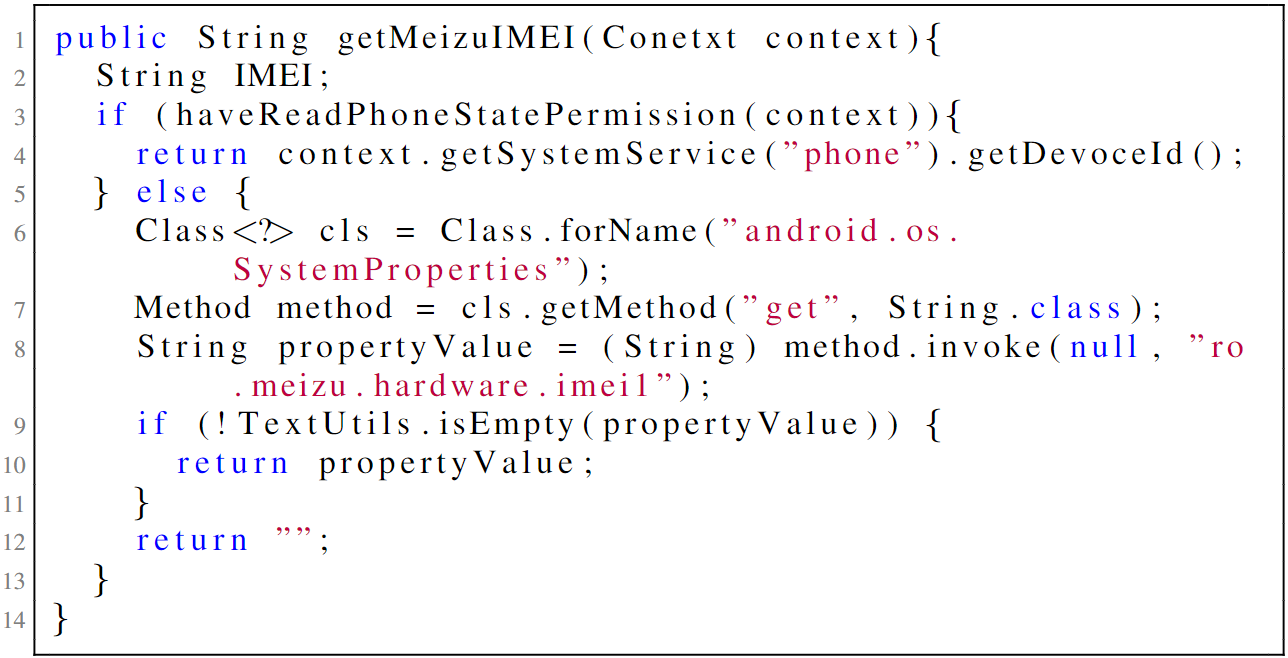}
  }
  \caption{Retrieving the IMEI of Meizu Devices through Customized System Properties.}
  \vspace{-0.1in}
  \label{fig:imei}
\end{figure}

\vspace{0.1em}
\noindent \textbf{System Properties Containing Hardware Identifiers.}
Manufacturers may accidentally introduce additional system properties that store sensitive information, potentially compromising user privacy.
These system properties, accessible without requiring any permissions by default, become potential targets for malicious exploitation.
We observed certain apps attempting to access the data stored in system properties introduced by the customized system, such as \texttt{\seqsplit{ro.meizu.hardware.imei1}}, \texttt{\seqsplit{ro.vendor.vivo.serialno}}, and \texttt{\seqsplit{ro.ril.miui.meid}}. 
Apparently, from their names, it appears that these properties contain user-unresettable hardware identifiers.
As demonstrated in the code snippet in \autoref{fig:imei}, this app manages to obtain the device's IMEI through the system property when the app is not granted permission to access IMEI (i.e., \texttt{\seqsplit{android.permission.READ\_PHONE\_STATE}}), infringing on user privacy.
We identified a total of 33 apps exhibiting similar behaviors.
Some apps are quite popular, e.g., ``com.global.wt'' gains 500K downloads on Google Play, ``com.wibo.bigbang.ocr'' have 2M downloads on Huawei AppGallery, and ``com.fengbo.live'' receives a significant 16 million downloads on Huawei AppGallery.

Furthermore, we attempted to verify these findings on real devices, utilizing cloud testing platforms~\cite{wetest, emas} to remotely conduct tests on various devices.
We developed an Android app that request no permissions and has the single functionality of retrieving all system properties from the device by executing the \texttt{\seqsplit{getprop}} command.
Notably, we identified the existence of system property \texttt{\seqsplit{ro.meizu.hardware.imei1}} on multiple Meizu devices (e.g., Meizu Note9 running Android 9, Meizu 16s Pro running Android 9 and Meizu 18s Pro running Android 11) and confirmed the system property can be accessed without any permissions.
In Android 9, such an exploit allows obtaining the device's IMEI without requiring permission or user consent. 
Even worse, despite Google's strict prohibition of third-party apps accessing the IMEI in Android 11, the exploitation of this customized system property enables third-party apps to freely retrieve the IMEI on devices from major manufacturers like Meizu, effectively circumventing the restrictions imposed by the Android platform.
In addition, we discovered that other hardware identifiers such as the MEID (Mobile Equipment Identifier) and the serial number, are also accessible without any permissions in the system properties of certain Meizu devices.
This circumvents Android's permission restrictions on apps, clearly infringing on user privacy.

Unfortunately, we were unable to confirm the exploitability of other system properties. 
In the cloud testing platform, we did not find any Vivo tablet devices, so we were unable to validate system property \texttt{\seqsplit{ro.vendor.vivo.serialno}}, which exists only in Vivo's tablet devices based on the identified device-specific behaviors.
On Xiaomi devices, we did find the existence of system properties storing hardware identifiers (i.e., \texttt{\seqsplit{ro.ril.miui.imei}} and \texttt{\seqsplit{ro.ril.miui.meid}}). 
However, due to access controls set by Xiaomi for these system properties, we cannot retrieve the values of these properties through our Android app.
It is worth noting that, since these system properties are stored in files, access control can be applied to them.
Only with higher-level permissions like ADB~\cite{adb}, we can access the content of these properties.

~\\
\noindent \fbox{
\parbox{0.45\textwidth}{
\textbf{Answer to RQ2:} 
\textit{Device-specific behaviors fall into three types: common compatibility issue fixes, manufacturer-introduced feature adaptations, and privacy-related behaviors.
Surprisingly, a considerable number of aggressive apps in the market abuse the customized system properties to get hardware identifiers without requiring any permissions, even some of them have been downloaded for millions of times. We advocate community to pay special attention to such kinds of behaviors.
}
}}

\subsection{RQ3: Sources of Device-specific Behaviors}

To investigate the origin of these device-specific behaviors, we initially extracted the package names of methods associated with these behaviors and recorded their frequencies.
The scrutiny of package names enabled us to distinguish whether these behaviors originated from SDKs or were embedded in the app's code by developers.
It is worth noting that we manually excluded obfuscated package names whose sources cannot be identified.
After this process, we obtained 879 distinct package names, and categorized the sources of behaviors into three types: well-adopted SDKs (\texttt{Type-I} SDKs), SDKs tailored to device-specific behaviors (\texttt{Type-II} SDKs), and developer-written code, as delineated in \autoref{tab:rq3}.

\texttt{Type-I} SDKs are famous SDKs (e.g., SDKs released by major companies ``com.facebook'' or push SDKs ``cn.jpush'') with a wide range of user base.
They include device-specific behaviors to ensure the proper functioning of the SDK across different devices.
As shown in \autoref{tab:rq3}, these SDKs primarily focus on device information collection across different customized systems, push services, and permission management.
\texttt{Type-II} SDKs are specifically designed to assist developers in implementing a specific functionality more reliably across devices.
Instances of Type-II SDKs encompass Agora SDK~\cite{agora}, designed for handling audio and video functionalities, ShortcutBadger SDK~\cite{badgesdk} for setting the number of unread message on home screen, and AutoStarter SDK~\cite{autostartsdk} for requesting self-start permission for the app.
These two types of SDKs constitute a majority of behavior sources, while the proportion of developer-implemented device-specific behaviors is relatively small (17\%).
We further discovered that in general, developers only manage to implement one or two types (an average of 1.8 types) of device-specific behaviors in their apps, which reflects the difficulty for developers to achieve such integration. 
Indeed, it is impractical for most developers to implement exhaustive device-specific behaviors across a wide range of device models and system versions.
This spurred our investigation into the availability and accessibility of related information from the developers' perspective to assess the challenge. \looseness=-1

\begin{table}[hbt]
\vspace{-0.1in}
\caption{The Distribution of Behavior Sources and Top 3 Prominent Behaviors Types for Each Source.}
\label{tab:rq3}
\resizebox{1\linewidth}{!}{
\begin{tabular}{|c|c|c|c|}
\hline
& \textbf{\texttt{Type-I} SDK}  & \textbf{\texttt{Type-II} SDK} & \textbf{Developer Code}  \\ \hline
Involved Apps & 1148 (49\%) & 1100 (47\%) & 411 (17\%) \\ \hline
\multirow{3}{*}{\begin{tabular}[c]{@{}c@{}}Top 3 Focused \\ Behavior Type\end{tabular}} & Device Information (75\%) & Audio (36\%)      & Status Bar (33\%) \\ \cline{2-4} & Push (70\%) & Launchers (29\%)  & Push (32\%) \\ \cline{2-4} 
 & Permission Management (50\%) & Status Bar (22\%) & Permission Management (22\%) \\ \hline
\end{tabular}
}
\vspace{-0.1in}
\end{table}

Our case study focuses on the official documentation provided by three major manufacturers, namely Samsung~\cite{samsungdev}, Xiaomi~\cite{xiaomidev}, and OPPO~\cite{oppodev}.
We browsed the official documentation and utilized its search functionality, attempting to find information about the 29 types of device-specific behaviors we identified.
Given the limited content in the global versions of Xiaomi and OPPO documentation, we referred to their Chinese documentation.
Among the \textbf{22} behaviors associated with Samsung devices, only information on \textbf{5} of them was found in Samsung's documentation.
Xiaomi's documentation covered \textbf{9/18}, while OPPO's official website covered \textbf{10/17}.
The available documentation predominantly highlights new features introduced by manufacturers, such as push services and foldables. 
Notably absent were details about compatibility issue fixes and comprehensive explanations of system customizations, such as modifications to screenshot storage locations or the addition of system properties.
These documents primarily serve to guide developers in understanding the new features within customized systems and how to effectively leverage them.
We further extended our search to GitHub and search engines.
Relevant information corresponding to our discoveries was located on technical forums like XDA Forums~\cite{xdaforums} and Stack Overflow~\cite{stackoverflow}, as well as on GitHub's issue descriptions and shared code snippets.
The only exception was the absence of information about custom system properties containing hardware identifiers.
In the retrieval process, we found that information related to device-specific behaviors is highly scattered. 
The keywords in \S\ref{sec:rulematching} were instrumental in assisting our search, while it may be challenging for developers to locate relevant information without any prior knowledge. 
We will further discuss how our work can assist developers in this process in \S\ref{sec:discussion}. \looseness=-1

~\\
\noindent \fbox{
\parbox{0.45\textwidth}{
\textbf{Answer to RQ3:} 
\textit{The source of device-specific behaviors in apps is predominantly associated with well-adopted SDKs and SDKs expressly designed to help developers implement such behaviors. 
Instances of developers independently coding these behaviors are relatively rare. 
Moreover, Information about device-specific behaviors is scant in manufacturers' official documentation; instead, this information is scattered online, posing challenges for developers acquiring relevant information.}
}}
\section{Discussion}
\label{sec:discussion}

\noindent\textbf{Implication.}
Our investigation revealed the widespread occurrence of device-specific behaviors within apps.
These behaviors play a crucial role in fixing compatibility issues, ensuring consistent performance across diverse devices, and capitalizing on novel features introduced by manufacturers to augment app functionality. 
It is noteworthy, however, that numerous apps exhibit no discernible device-specific behaviors, potentially leading to operational irregularities or the absence of certain features on devices.
Some developers might lack awareness of specific issues or features tied to particular devices, or they may be uncertain about how to deal with them. 
Our research aims to equip developers with a deeper comprehension of device-specific behaviors, delineating potential issues in app functionalities across different devices and offering guidance on developing corresponding device-specific behaviors. 

Through examination, we noticed that manufacturers' official documentation mainly emphasizes newly introduced features, overlooking comprehensive information about system-specific issues and alterations to existing AOSP functionalities.
We advocate for a more proactive stance from manufacturers in furnishing detailed insights into device-specific issues and their respective solutions, and providing necessary clarifications about system customization, such as permission management.

Furthermore, our result revealed that certain system customizations, such as the system properties containing hardware identifiers, can pose security risks. 
The security issues detected within apps may only represent the tip of the iceberg. 
Manufacturers must prioritize system security during customization, especially concerning privacy-related changes, where sound authentication methods are crucial.

\noindent\textbf{Limitation.}
First, our approach involved static analysis to examine device-specific behaviors in large-scale real-world apps, focusing on Java code within unpacked apps.
This prevented us from addressing packed apps, native code, and dynamically loaded code.
Additionally, due to limitations in static analysis tools, some apps couldn't be analyzed despite being unpacked.
To validate system properties containing hardware identifiers, we utilized devices from cloud testing platforms, as we lacked access to specific devices. 
However, it's uncertain if the platform's customizations to these devices for cloud testing could affect their security and consequently impact our experimental results.
\section{Related Work}
\label{sec:related}

\noindent \textbf{API-related Compatibility Issues.}
API-related compatibility issues have been extensively researched. 
Li et al.~\cite{li2018cid} developed a method to identify these issues directly from Android app bytecode using static analysis techniques. 
They modeled API lifecycles to assess whether the app's API usage might cause compatibility problems.
Huang et al.~\cite{huang2018understanding} investigated callback API issues, creating the CIDER tool for detection via static analysis and employing Callback Control Flow Graphs for inconsistency identification. Xia et al.~\cite{xia2020android} presented RAPID, a tool for automatically detecting addressed compatibility issues in apps, and classified compatibility checks using a learning-based approach.

\noindent \textbf{Device-related Compatibility Issues.}
Compared to API-related compatibility issues, research on device-related compatibility issues is not as comprehensive.
Zhao et al.~\cite{zhao2022towards} developed RepairDroid, a prototype tool capable of automatically addressing compatibility issues in Android apps. 
They introduced an app patch description language for articulating solutions to various compatibility issues, including device-related ones.
Wei et al.~\cite{wei2016taming} manually inspected the source code of 27 open-source apps to identify device-related compatibility issues. They also proposed a formalism for modeling these issues and created FicFinder for automated detection. However, FicFinder focuses solely on open-source apps, which may lack the rigorous device-specific refinement process seen in commercial app market apps, as noted in previous studies~\cite{liu2023towards}.
Wei et al.~\cite{wei2019pivot} introduced Pivot, a technique for automatically extracting API-device correlations from Android apps and prioritizing them based on their likelihood of causing compatibility issues. They validated Pivot's effectiveness using over 5,000 apps from Google Play and identified 17 compatibility issues, although only scattered ones corresponded to 10 functionalities.
\section{conclusion}
We performed an in-depth examination of device-specific behaviors in a broad range of real-world Android apps.
Our extensive analysis, covering over 20,000 apps, identified 2,357 apps exhibiting device-specific behaviors. 
These behaviors were primarily employed for issue fix and adapting to features introduced by manufacturers. 
Furthermore, we uncovered device-specific behaviors related to privacy. 
Certain apps leverage unprotected system properties to extract non-resettable identifiers without the need for any permissions, posing a substantial risk to user privacy.
Our research not only offers developers valuable insights into device-specific behaviors but also underscores the imperative for enhancing documentation and fortifying the security of customized code.

\newpage
\bibliographystyle{IEEEtran}
\bibliography{base}

\end{document}